\documentclass{elsart}

\usepackage{graphicx}
\usepackage{amsmath}
\usepackage{amsfonts}
\usepackage{amssymb}
\usepackage{color,graphicx,shortvrb,epsfig}
\usepackage{latexsym}

\newcommand{\gsim}{\lower.7ex\hbox{$\;\stackrel{\textstyle>}{\sim}\;$}}
\newcommand{\lsim}{\lower.7ex\hbox{$\;\stackrel{\textstyle<}{\sim}\;$}}
\newcommand{\tturn}{T_{\rm parabola}}

\begin{document}
\begin{frontmatter}
\title{Detecting neutrino transients with optical follow-up observations}

\author[HU]{Marek Kowalski\corauthref{cor}}
\corauth[cor]{Corresponding author.}
\author[HU]{Anna Mohr}
\address[HU]{ Humboldt University, 12489 Berlin, Germany} 

\date{15. March 2007}

\begin{abstract}
	 A novel method is presented which  will enhance the
sensitivity of neutrino telescopes to identify transient sources such as 
Gamma-Ray Bursts (GRBs) and core-collapse Supernovae (SNe). Triggered by 
the detection of high energy neutrino events from  IceCube or other large 
scale neutrino telescopes, an optical follow-up program will allow the 
identification of the transient neutrino source. We show that once the 
follow-up program is implemented, the achievable sensitivity of IceCube to 
neutrinos from SNe and GRBs would increase by a factor of 2-3. The program 
can be realized with a small network of automated 1-2 meter telescopes and 
has rather modest observing time requirements.
\end{abstract}
  

\end{frontmatter}


\section{Introduction}
IceCube, the first instrumented gigaton detector being constructed, will have
an unprecedented sensitivity for neutrinos of TeV to PeV energies \cite{Ahrens:2003ix}. 
The list of 
astrophysical objects which might be detected is extensive (see \cite{Halzen:2002pg} for a review)  and includes 
time variable sources such as Active Galactic Nuclei (AGNs) and Micro-quasars  
as well as transient sources such as Gamma-Ray Bursts (GRBs) and 
Supernovae (SNe). Because the number of expected events is small, 
 efficient search strategies need to be developed to separate the signal 
from the 
background of atmospheric neutrinos. 
Point sources are identified through the direction of muon-neutrino events 
which can be reconstructed to 
within $\sim1^\circ$ \cite{Ahrens:2003ix}. 
For time variable sources
one can additionally  search for a 
correlation of neutrino arrival times with the activity level of the
source observed e.g.\ in gamma-ray, X-ray \cite{elisas} or optical 
\cite{wiyn}. Although  promising, 
this approach is made more complicated by the fact that the correlation 
features as well as their time-scales and rates are a priori not well known. 
This is somewhat simpler in case of transient sources. For GRBs, 
the time constraints from the burst detected by satellites allows one to 
eliminate essentially all atmospheric neutrino background \cite{amanda_grb}.

A practical problem  is that most gamma-ray, X-ray 
and optical observatories are capable of observing only a small fraction 
of the 
sky, while neutrino telescopes monitor essentially a full hemisphere. Hence, 
unless preparatory steps are taken, only a small subset of data can be used 
for correlation studies. 
The solution we discuss in this paper consists of using 
neutrino detections to trigger other follow-up observations.
As we show,  such a Target-of-Opportunity (ToO) program can provide the 
required data to identify transient sources (see \cite{elisas} for a discussion of this idea in the context of AGNs).  
As examples we consider two promising source classes, namely  
GRBs and SNe. 
  
GRBs are among the most spectacular events in the 
Universe, releasing $\sim10^{51}$~ergs on time-scales of seconds. 
These observations, which are best explained by the presence of 
highly relativistic jets with boost factors $\Gamma\sim100$, 
have fueled speculations 
that GRBs might be the source of the most energetic cosmic rays
\cite{waxman95,vietri95}.
Protons would be accelerated to high energies, and in the interaction with 
ambient photons produce pions and kaons, which in their decay
produce neutrinos  \cite{wb97,wb99,lategrb,grb_fluk}. 

Furthermore, the recent association of Gamma-Ray Bursts with core-collapse 
Type Ib/c
Supernovae might  suggest that certain aspects of the jet 
phenomenon are present in other, more frequent types of SNe. 
(The ratio of the local GRB rate, including a jet beaming correction, 
to the rate of SNe Ib/c 
was found to be less than $10^{-3}$ \cite{rate}.)
In particular, jets might form with very different
ejection velocities, some highly relativistic leading to the GRB phenomenon, 
while others being only mildly relativistic. A mildly relativistic jet ($\Gamma$ of a few) would stall in
the outer layers of the progenitor star resulting in the absorption of 
all its electromagnetic radiation. 
However, such  a jet would efficiently
sweep up matter and accelerate protons. Inevitably, these protons
interact with other surrounding nuclei, 
producing neutrinos. The predicted neutrino
yield from the interaction of jets with
the outer layers of the
progenitor  has been calculated by several authors
\cite{Meszaros:2001ms, Razzaque:2004yv} and recently
shown to be a potentially very strong signal
\cite{Ando:2005xi,Razzaque:2005bh}.

The observational program discussed in this paper consists of 
a ToO trigger and optical follow-up observations. As we will show, this
would improve the perspectives for discovery of neutrinos from 
transient sources significantly. We quantify the gain in sensitivity
for neutrinos from GRBs and SNe and discuss the telescope and observing 
time requirements.



\section{Neutrino Trigger}
\label{sec:nu_trigger}
The difficulty in triggering a ToO follow-up program lies in the fact that the 
neutrino data stream is dominated by atmospheric
neutrinos. We first 
consider the rate of atmospheric background events in a detector such as 
IceCube. Near the energy threshold of the detector, a detailed simulation of 
the
detector performance is necessary, resulting in an 
event rate from atmospheric muon neutrino events of $9\times10^4$ per year 
(after application of neutrino selection cuts) \cite{Ahrens:2003ix}. 

Above a muon energy threshold of 10 (100)~TeV, one
expects of the order of 2500 (80) events per year from atmospheric muon 
neutrinos \cite{marek05} (assuming a muon 
energy resolution of  $\sigma(\log E_\mu)=0.4$). In the case of GRBs, 
the expected neutrino spectrum is significantly harder than 
that of atmospheric neutrinos, hence one can introduce an energy cut to 
efficiently reduce the background.

When searching for transient events such as SNe or GRBs, one obvious 
signature to look for are neutrino-bursts: a multiplet 
of neutrino events from the same direction and 
within a short time window. 
Let's consider the background rate from $n$ coincident, uncorrelated, upwards 
traveling atmospheric neutrinos:
\begin{equation}
  R^{\rm atmo}_{n-\nu} \approx n!\left[\frac{\Delta \Omega}
{2\pi} \Delta t\right]^{n-1} (R^{\rm atmo}_{1-\nu})^n.
\label{eq:multi}
\end{equation}

Inserting the corresponding values for IceCube, an angular 
accuracy of  $\Delta\Omega$=$2^\circ\times2^\circ$ and a
rate, $R^{\rm atmo}_{1-\nu}$, of $9\times10^4$ atmospheric neutrino events 
per year as well as a
  duration of the neutrino-burst $\Delta t$ of 100 seconds, one can
 estimate the background rate for higher order multiplets.
A doublet ($n=2$) is expected
at a rate\footnote{the rate of background multiplets is very sensitive 
    to the rate of atmospheric neutrinos and the angular bin-size.
While it is possible to exactly compute the rate for any experimental setup,  
it will generally differ somewhat from the quoted value.}  of 10 ${\rm year}^{-1}$.  Hence, the 
detection of a doublet would by itself
not be significant. As will be 
discussed below, this changes
once an optical transient event such as a GRB or a SN is detected in 
coincidence. If the 
neutrino multiplet consists of more than two events, it becomes
significant. The $2 \times 10^{-3}~{\rm year}^{-1}$ rate of triple coincidence  
background atmospheric neutrino events poses little threat of 
false trigger.

\section{Follow-up observations}
\label{sec:optical_rec}
Searches performed with optical 
telescopes have been shown to be by far the most efficient in detecting
supernovae.  Depending 
on the supernova type, the light-curve brightens for about 10-20 days after 
explosion before reaching its maximum. If the rising part of the
light-curve is measured well enough,
one can extrapolate 
to the explosion time $t_0$. For GRBs, one would rely on detection of the 
afterglow, which has been observed for about $50\%$ of the well localized
GRBs \cite{piran05}. The measurement of the afterglow 
can be extrapolated to obtain an estimate for $t_0$. We show that for both 
SNe and GRBs, a resolution $\Delta t_0 \lsim 1$~day can be achieved with a 
rather modest optical follow-up program.

\begin{figure}
\begin{center}
\includegraphics[width=0.4\textwidth]{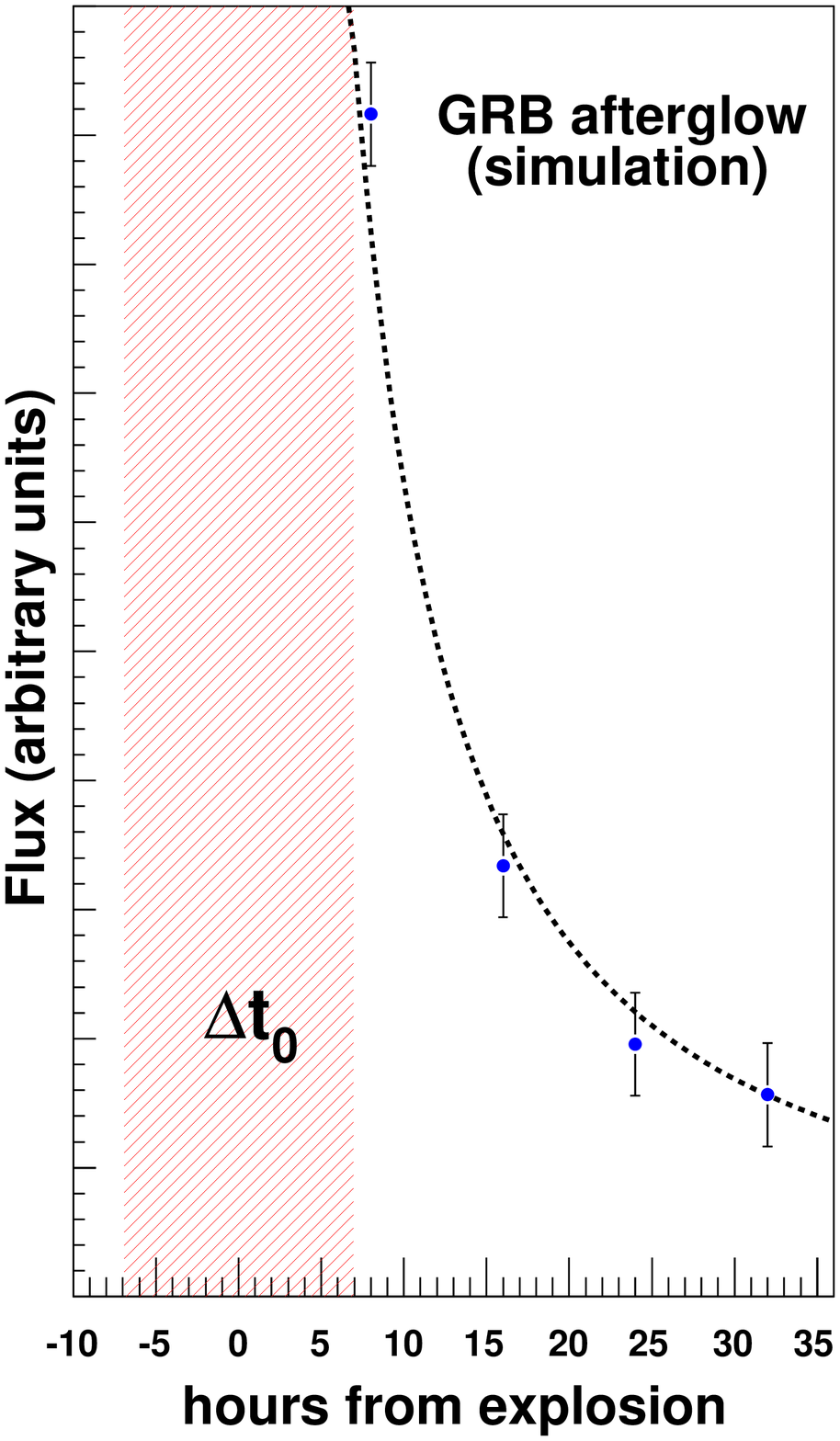}
\hfill
\includegraphics[width=0.4\textwidth]{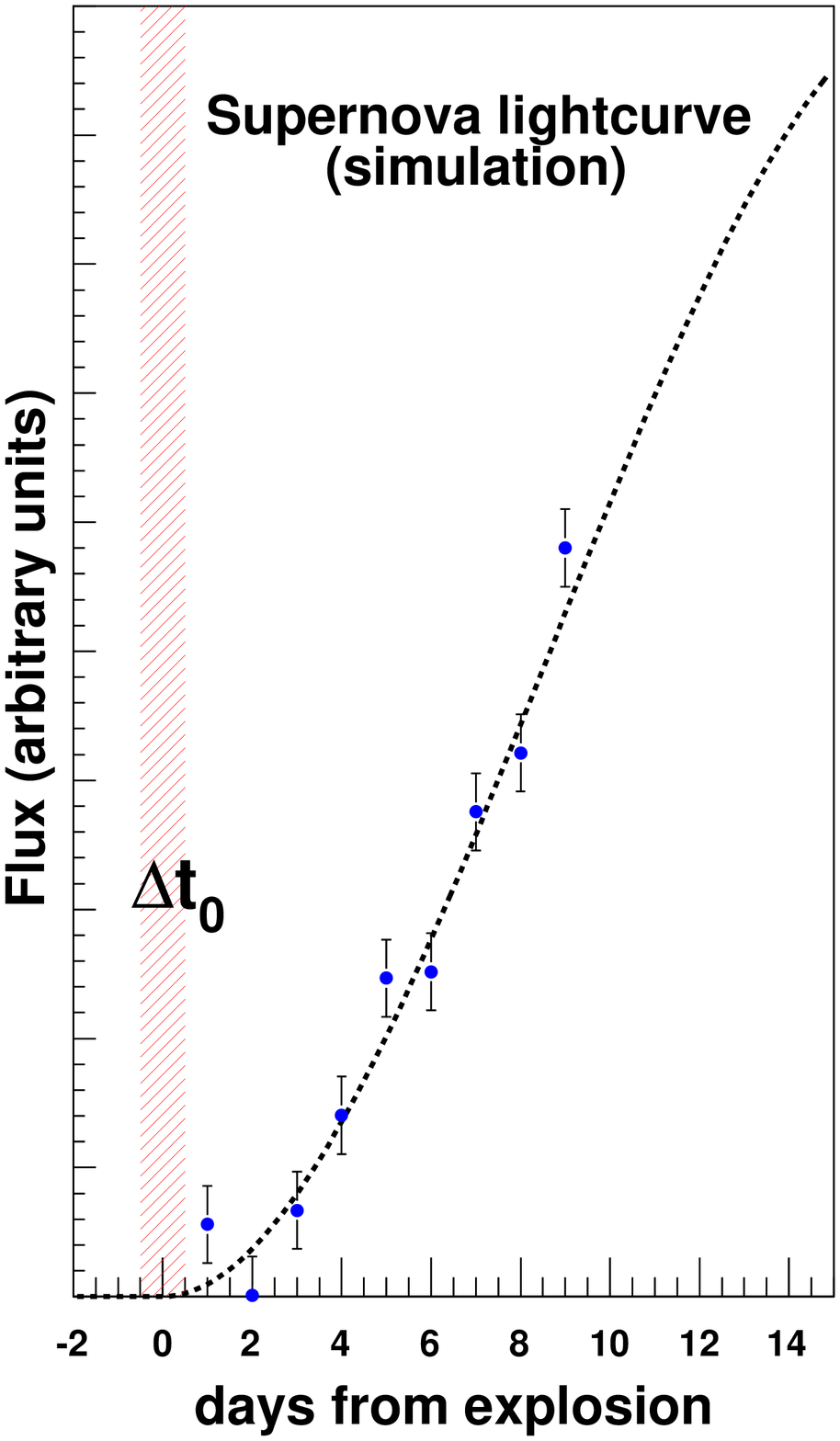}
\caption{Schematic determination of $t_0$ from a GRB afterglow (left) and supernova lightcurve (right). The data was sampled from the model (dotted line)
with a flux uncertainty of $\sim5\%$ for the brightest 
data point. The
shaded region represents the average achievable uncertainty for $t_0$. 
\label{fig:ltcv}}
\end{center}
\end{figure}

{\it Gamma-Ray Bursts:}
The afterglow 
decays initially as a power-law $f(t)=a (t-t_0)^{-\gamma}$, with an average 
index  of $\gamma \sim 1.2$ \cite{piran05}. One day after the burst, 
a typical brightness of the afterglow is 19-20 magnitudes.
Such afterglows therefore represent a promising target for a neutrino triggered
optical follow-up. To determine with which precision one can obtain the
explosion date $t_0$, we simulate the afterglow with a power-law index $\gamma=1.2$ and fit for the
parameters of the light-curve model: $a,\gamma$ and $t_0$. 
With three observations of the afterglow at 8, 16 and 24 hours after the 
burst, and the first being a 20 
sigma detection (i.e.\ an error on the flux of 5\%)
, we obtain an average time 
resolution  $\delta t_0\sim0.3$~days (see Fig.\ \ref{fig:ltcv}).

{\it Supernovae:}
The explosion time can be obtained by extrapolating the supernova light-curve.
A simple model for the initial, rising part of the light-curve is considered. We assume that, in analogy to Type Ia supernovae,
the lightcurve can be approximated by
 $F\propto(t-t_0)^2$ \cite{conley06}
(This behavior can be understood by 
assuming that in the initial phase after the explosion, the 
supernova photo-sphere is represented by a black body of constant temperature, 
 which expands with velocity $v$. The area of the 
photo-sphere, which is directly proportional to the photon flux, 
then increases $\propto v(t-t_0)^2$.) 
The time after 
which the light-curve evolution begins to slow down, and start to deviate 
from  the $t^2$-parameterization, will be called
 $\tturn$. 
For Type Ia Supernova, one finds $\tturn \sim 10$~days. We will 
assume that a similar relation holds for core-collapse supernovae, however with 
different $\tturn$. Generally, Type II supernovae have a very 
fast rise-time corresponding to a short  $\tturn$ of a few days while Type Ib/c have a slower rise-time.  
We will make the general ansatz for the early time flux evolution:
$
F(t)=a(t-t_0)^2 
$
for $ 0<(t-t_0)< \tturn $.
This defines a  template for times before $\tturn$ which, in combination 
with the observed lightcurve, can be used to constrain the parameters $a$ and $t_0$.  
An observational program with daily observations of the 
same field is assumed. The exposure time is chosen such that for the range of target supernovae, the flux error at the time $\tturn$ is $\sim5~\%$ (see Fig.\ \ref{fig:ltcv}). One 
can then estimate the achievable measurement errors from the light-curve up 
to times $\tturn$ using simulation.
We find that the achievable statistical error on  $\Delta t_0$  is about 
half a 
day, and largely independent of $\tturn$ (and hence the rise-time). Even if we 
neglect the information of the 
first three days (which might be showing the shock breakout), the expected 
error increases only  to $\Delta t_0\approx0.8$  days.

In cases where the shock breakout can be detected directly, the supernova is 
seen as an initially bright object, with a rapidly cooling and declining 
light-curve. A well observed Type Ib/c supernova showing  a shock breakout
is  SN 1999ex \cite{1999ex}. 
The determination of $t_0$ can be done in a manner similar to the case of GRB afterglows.

\section{Neutrino-Optical Coincidences}
\label{sec:combined}
After introducing all the necessary ingredients in the previous sections, 
we now turn to the strategy for a combined neutrino/optical detection. The 
basic scheme is that the detection of neutrino-induced muons triggers 
the optical observations. Since supernovae and GRBs differ in their rate and 
in their expected neutrino spectra, they need slightly different strategies.

{\noindent \it Supernovae:}
Three trigger scenarios are considered.

{\noindent \it 1) Neutrino-triplets and higher order multiplets}: From Eq.\ \ref{eq:multi} follows that
three or more neutrinos detected within a small angular and temporal
window constitute a statistically significant neutrino-burst. Optical follow-up would then serve the purpose of identifying
the source of the neutrino-burst.

{\noindent \it 2) Neutrino-doublet}: A neutrino-doublet in IceCube 
is expected several times a year from the background of atmospheric neutrinos.
A doublet detection 
by itself is hence not significant. However, this changes 
if a coincident supernova is detected. The rate of core-collapse supernovae 
is about
one per year within a 10 Mpc sphere \cite{snerate}. Thus, within a
time-window of one day, an angular window of
$\Delta\Omega =2^\circ\times2^\circ$ and within a
distance $d_{\rm max}$, we expect to observe a supernova with a probability $P^{\rm bg}_{\rm SN} \approx 3 \cdot 10^{-7}\times (d_{\rm max}/\mbox{10 Mpc})^3$.
For $d_{\rm max}$ =200~Mpc, the expectation to observe a background 
doublet in coincidence with a random supernova is $R^{\rm atmo}_{2-\nu} P^{\rm bg}_{\rm SN}(d<200~\rm Mpc)\approx2 \cdot 10^{-2}$. If in addition one would require that the supernova is of Type Ib/c, the background expectation 
rate would reduce by another factor of 6 \cite{rate_1bc} to $3 \cdot 10^{-3}$ per year.
With such a small background expectation, a neutrino-optical coincidence would 
become a very interesting detection.
Finding supernovae with $d_{\rm max}\gg 200$ Mpc can only be associated
with small confidence (unless they have some features making them exceptional supernova, as for example a Gamma-Ray Burst afterglow). 

\begin{figure}
\begin{center}
\includegraphics[width=0.8 \textwidth]{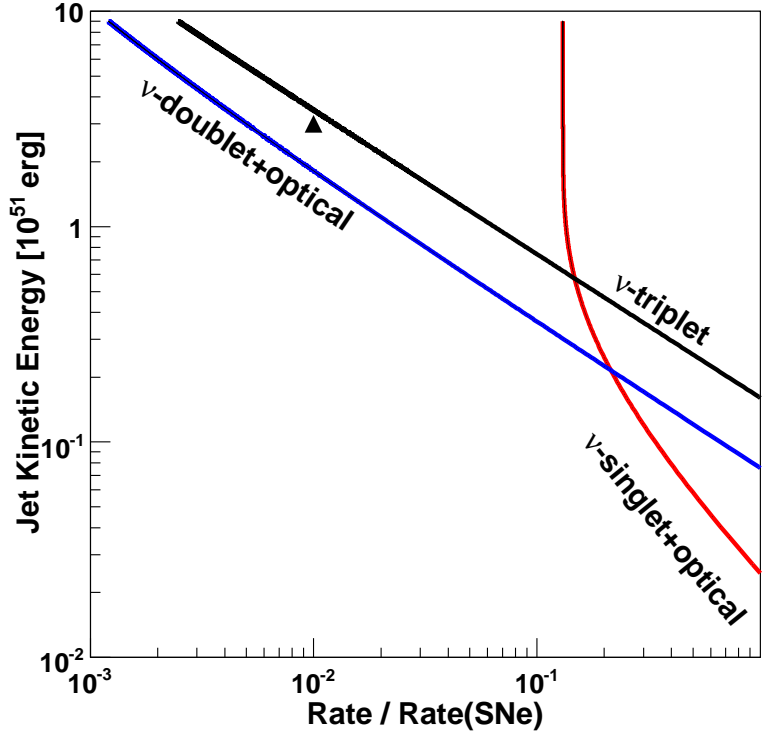}
\caption{Limiting sensitivity on the jet kinetic energy as a function of the rate of jet producing supernova normalized to the rate of SNe of all types. The three different lines represent the one-detection-a-year contours for 
neutrino singlets, doublets and triplets as 
discussed in the text. The triangle represents the prediction of \cite{Ando:2005xi}.}
\label{fig:exclusion}
\end{center}
\end{figure}

{\noindent \it 3) Neutrino-singlet}: Because of their large rate, single neutrinos are rather difficult to associate with a supernova. The maximal distance allowed 
has to be restricted accordingly. For example, for a rate $R^{\rm atmo}_{1-\nu}$ of $9\times10^4$ neutrino events per year and $d_{\rm max}=20$~Mpc,
 one obtains a rate of accidental coincidences of $R^{\rm atmo}_{1-\nu} P^{\rm bg}_{\rm SN}\approx0.2$ per year.
At least three or four such coincidences have to be detected to make 
it meaningful (i.e.\ $\gsim3\sigma$). Because the number of galaxies within 
20~Mpc is still reasonably small, a 
nightly scan of a catalog of galaxies would still be feasible 
(see \cite{nosweat} for an ongoing program to monitor the largest 
galaxies within 10~Mpc).

We can now compute the sensitivities of the three observing scenarios 
described above. We assume a model with two parameters. The first parameter is 
the rate of supernovae producing internal jets and the second parameter is the kinetic energy released into the jet.
We scale the expectation from the model
of \cite{Ando:2005xi}, 30 events above a muon energy of 100~GeV for a detector of km$^2$ area,
with the factor $(E_{\rm jet}/3\cdot10^{51} \mbox{ erg})$.
We distribute the supernovae according the continuum distribution of 
\cite{snerate}, and weight their contribution with the distant dependent 
probability to produce one, 
two, three or more neutrino events in the detector. 
The inclusion of Poisson-fluctuations in the number of 
expected events leads to a large increase in the number of detectable 
supernovae, as one samples from a larger volume. For example, the median 
distance of doublet producing supernovae is 120~Mpc, compared to 40~Mpc 
without Poisson fluctuations.  

We have taken into account that about 
20\% of supernovae (and GRBs) cannot be followed optically 
because they occur to close too the sun. We do not include the impact of dust 
extinction, which will hide a fraction of SNe in starburst galaxies 
from their detection with optical follow-up \cite{snerate,dust}. This 
fraction will depend on the 
wavelength range of observations and could be somewhat reduced by observing in 
the near infrared or by more powerful telescopes.

Figure \ref{fig:exclusion} shows the achievable constraints on 
the jet energy as a function of the rate of supernovae with jets. 
As can be seen, neutrino-doublets, if combined with an optical follow-up can improve the sensitivity by a factor of a few,
either by lowering the accessible
supernova rate or by lowering the energy of jets, which can still be detected.
 The 
constraints from neutrino-singlets show a cut-off around a fraction of  $\sim0.1$ which is a consequence of having to restrict the distance of the supernova to be within 20~Mpc. Hence, rare supernovae jet configuration can not be efficiently 
probed with neutrino singlets. 

{\it Gamma-Ray Burst:}
In order to compute the reach of a neutrino-optical coincidence search, 
we first estimate the rate of accidental afterglow detections consistent 
with the timing and direction of a neutrino event. Assuming 500 GRBs per year 
and hemisphere,  
of which 
$50~\%$ produce detectable afterglows \cite{piran05}, 
a coincidence time window $\delta t_0\sim0.3$~days and an 
angular resolution of $1^\circ$, the probability
for an accidental 
coincidence is $\sim3 \cdot 10^{-5}$ for each detected neutrino event. 
Hence, in order to make a meaningful 
detection, the rate of neutrino triggers should not exceed several hundred.
 The neutrino energy spectra expected 
from GRBs are significantly harder than that of atmospheric neutrinos. By  
demanding that the reconstructed muon energy is higher than a certain 
threshold energy, the number of atmospheric neutrino events can be 
reduced significantly.
We have extended the 
simulation of \cite{marek05} to include the 
GRB neutrino flux prediction of 
\cite{wb99}. Assuming a muon energy resolution of $\sigma(\log E_\mu)=0.4$, we 
find that about 50\% (16\%) of neutrino events would have a reconstructed 
energy larger than 10 (100)~TeV. 
Hence, 
taking into account the afterglow and neutrino detection probability, as well the fact that about 20\% of the sky can not be observed due to the sun, 
every 5th (16th) GRB neutrino event could be identified as such, 
with about 93\% (99.98\%) confidence.
With an 
expectation of about 10 events from GRB per year and hemisphere \cite{wb99},  
this would lead 
to up to 2 GRB detections per year.

Note that for GRBs, unlike SNe, the rate is so low and the luminosity so
high,
that one samples throughout cosmological distances. The variations
due to GRB distances have less dramatic consequences  and as a result,
the ratio
of doublets to singlet neutrino detections will be smaller for GRBs.
We will consider one specific calculation as an example. 
In \cite{grb_fluk}, individual neutrino
 event rates where modeled for 579 long duration GRBs. For Model I, they 
obtain about 6 muon neutrino events per 1000 bursts for a cubic kilometer 
array. The expected rate of doublets for this sample is 0.3. Hence, taking 
into account the efficiencies discussed above, 
the rate of identifiable doublets will be observed at a rate a few times lower 
than singlets.

\section{Discussions and Conclusions}
As illustrated here, an optical follow-up of interesting events, 
either neutrinos of very high energy or neutrino multiplets ($n\geq2$), would 
improve the perspective for neutrino detection from SNe and GRBs significantly. 

The follow-up 
can be achieved with a rather modest observing program.
Depending on the neutrino-trigger settings, a ToO would be issued 
several times a year (multiplet detections) up to several times a day 
(events with $E_\mu\geq 10$~TeV).
In case of a detection of a transient object,
lightcurves are constructed from repeated observations every few hours during 
the first night (to measure the afterglow) and then once a night for the next 
10 days. Optical 
telescope with 1-2 meter apertures can acquire the 
needed signal-to-noise (5\% flux error for a 20th magnitude point-source) 
within about one minute exposure time. 

To reduce the number of pointings,  the 
field of view (FoV) of the follow-up telescope should ideally  match 
the resolution of the neutrino telescopes ($\sim 1^\circ$ for IceCube \cite{Ahrens:2003ix} and less than half that for large-scale water cherenkov detectors \cite{km3net}). Such optical telescopes/instruments
already exist and many more are being planned or constructed. For example, 
the ROTSE III network consists of four 0.45 meter 
robotic telescopes with a $1.85^\circ \times 1.85^\circ$ FoV, which has been successfully 
operated since 2003. The MegaCAM camera on the 
CFHT 3.6 meter telescope, which is also in operation since 2003, 
has a FoV of $1^\circ \times 1^\circ$. With these 
telescopes, searches for orphan afterglows have already been performed 
and it was shown that the astrophysical background, e.g. due to variable stars, can 
be fully controlled \cite{rotse,cfht}.

For supernovae, the detection rate of neutrino transients would 
improve by a factor of $\sim3$ compared to the case without follow-up. For
 GRBs, which are sampled over large cosmological distance, the advantage might 
be even larger.
In particular, the afterglow follow-up observations proposed here could 
become an important counterpart to the search for a correlation between a 
neutrino signal and gamma-ray bursts. Dedicated 
satellites, such as SWIFT \cite{swift_bat}, are very efficient in 
finding GRBs and allow for a 
reduction of the coincidence search time-window to the duration 
of the gamma-ray 
burst. However, limited sky coverage results in a reduced detection rate. 
With a FoV of 1.4~sr, the SWIFT satellite triggers on only every 
9th GRB, 
while the neutrino-triggered optical follow-up, with the right
trigger-settings, allows the detection of about 
every 5th neutrino producing GRB. Finally, there is the possibility of 
neutrino production in GRB afterglow flares \cite{lategrb}. Due to the effect 
of jet broadening,  such GRBs might be more efficiently detected through their 
afterglow observations.
   
Perhaps as important as the gain in sensitivity is the acquired 
ability to identify the transient source of a neutrino burst, once it is 
detected with neutrinos. In this work we were guided by the phenomenology of  
GRBs and SNe. Yet, the optical lightcurves obtained from the 
follow-up observations of neutrino events might equally 
well lead to the discovery of other transient or variable 
sources of high-energy neutrinos.

{\it Acknowledgment:} We would like to thank John Beacom, Doug Cowen and Soeb Razzaque 
for valuable discussion. This work has been 
supported by the Deutsche Forschungsgemeinschaft (DFG).


\begin{thebibliography}{1}
\bibitem{Ahrens:2003ix}
  J.~Ahrens {\it et al.},
  Astropart.\ Phys.\  {\bf 20} (2004) 507.



\bibitem{Halzen:2002pg}
  F.~Halzen and D.~Hooper,
  Rept.\ Prog.\ Phys.\  {\bf 65} (2002) 1025.

\bibitem{elisas}E. Bernardini {\it et al.} 2005, Proc. of Workshop {\it Towards a Network of Atmospheric Cherenkov 
Detectors}, Palaiseau, France (2005), arXiv:astro-ph/0509396, E. Resconi, 
Proc. of {\it TeV Particle Astrophysics}, Madison, USA (2006).
\bibitem{wiyn} M. Bayer et al., Proc. of {\it TeV Particle Astrophysics}, Madison, USA (2006). 


\bibitem{amanda_grb}K.\~Kuehn et al., in Proc $29^{th}$ {\it Int. Cosmic Ray Conf.} (2005); M.\~Stamatikos ibid; B.Hughey, ibid; (see astro-ph/0509330).

\bibitem{waxman95}E. Waxman, Phys. Rev. Lett. 75  (1995) 386.

\bibitem{vietri95}M. Vietri, Astrophys. J., 883  (1995). 

\bibitem{wb97}
  E.~Waxman and J.~N.~Bahcall,
  Phys.\ Rev.\ Lett.\  {\bf 78}  (1997) 2292.

\bibitem{wb99}
  E.~Waxman and J.~N.~Bahcall,
  Phys.\ Rev.\ D {\bf 59}  (1999) 023002.

\bibitem{lategrb} K.\~Murase and S.\~Nagataki, Phys.\ Rev.\ Lett.\ {\bf 97}
 (2006) 051101.

\bibitem{grb_fluk}D.Guetta, D. Hooper, J. Alvarez-Muniz, F. Halzen and E. Reuveni, Astropart. Phys. {\bf 20}, (2004) 429; http://www.arcetri.astro.it/~dafne/grb/





\bibitem{rate} T.~Le, \& C.D.~Dermer\ 
(2006), arXiv:astro-ph/0610043. 

\bibitem{Meszaros:2001ms}
  P.~Meszaros and E.~Waxman,
  Phys.\ Rev.\ Lett.\  {\bf 87} (2001) 171102. 



\bibitem{Razzaque:2004yv}
  S.~Razzaque, P.~Meszaros and E.~Waxman,
  Phys.\ Rev.\ Lett.\  {\bf 93} (2004) 181101.
  [Erratum-ibid.\  {\bf 94} (2005) 109903]; S.~Razzaque, P.~Meszaros and E.~Waxman, Phys. Rev. {\bf D 68} (2003) 083001.



\bibitem{Ando:2005xi}
  S.~Ando and J.~F.~Beacom,
  Phys.\ Rev.\ Lett.\  {\bf 95} (2005) 061103.

\bibitem{Razzaque:2005bh}
  S.~Razzaque, P.~Meszaros and E.~Waxman,
  Mod.\ Phys.\ Lett.\ A {\bf 20} (2005) 2351.


\bibitem{marek05}
  M.~Kowalski,
  JCAP {\bf 0505}  (2005) 010.

\bibitem{piran05}
  T.~Piran,
  Rev.\ Mod.\ Phys.\  {\bf 76}  (2004) 1143.


\bibitem{conley06}A.~Conley, et al., Astrophys.J., 132 (2006) 1707. 

\bibitem{1999ex} M.~Stritzinger et 
al., Astrophys.J., {\bf 124}  (2002) 2100.


\bibitem{snerate}
  S.~Ando, J.~F.~Beacom and H.~Yuksel,
  Phys.\ Rev.\ Lett.\  {\bf 95} (2005) 171101. 

\bibitem{rate_1bc}
  E.~Cappellaro, R.~Evans and M.~Turatto,
  Astron.\ Astrophys.\  {\bf 351} (1999) 459.


\bibitem{nosweat}A.~Gal-Yam {\it et al.}, http://www.astro.caltech.edu/$\sim$avishay/nosweat.html

\bibitem{dust}
  F.~Mannucci et al.\,
  Astron.\ Astrophys.\  {\bf 401}  (2003) 519.




\bibitem{km3net}  Y.~Becherini {\it et al.},
  Nucl.\ Instrum.\ Meth.\  A {\bf 567} (2006) 477;
  U.~F.~Katz,
  Nucl.\ Instrum.\ Meth.\  A {\bf 567} (2006) 457. 
\bibitem{rotse}E.~S.~Rykoff {\it et al.},
  Astrophys.\ J.\  {\bf 631} (2005) 1032.

\bibitem{cfht}F.Malacrino et al.\, (2007), astro-ph/0701722.

\bibitem{swift_bat}S.~D. Barthelmy, et 
al.\, Space Science Reviews  120 (2005) 143.


\end{thebibliography}
\end{document}